\documentstyle [12pt,epsf]{article}

\newcommand{\beq}{\begin{equation}}
\newcommand{\dd}{\partial}
\newcommand{\eeq}{\end{equation}}
\newcommand{\bea}{\begin{eqnarray}}
\newcommand{\eea}{\end{eqnarray}}

\newcommand{\vf}{\varphi}

\newcommand{\e}{{\cal E}_\omega}

\begin{document}
\baselineskip 7.3 mm

\def\thefootnote{\fnsymbol{footnote}}

\begin{flushright}
\begin{tabular}{l}
CERN-TH/97-69 \\
hep-th/9704073 \\
April, 1997 
\end{tabular}
\end{flushright}

\vspace{2mm}

\begin{center}

{\Large \bf 
Small Q balls 
}
\\ 
\vspace{8mm}

\setcounter{footnote}{0}

Alexander Kusenko\footnote{ email address:
kusenko@mail.cern.ch} \\
Theory Division, CERN, CH-1211 Geneva 23, Switzerland \\

\vspace{12mm}

{\bf Abstract}
\end{center}

We develop an adequate description of 
non-topological solitons with a small charge, for which the thin-wall
approximation is not valid.   There is no classical lower limit on the
charge of a stable Q-ball.  We examine the parameters of these small-charge
solitons and discuss the limits of applicability of the semilcassical
approximation. 

\vfill

\pagestyle{empty}

\pagebreak

\pagestyle{plain}
\pagenumbering{arabic}
\renewcommand{\thefootnote}{\arabic{footnote}}
\setcounter{footnote}{0}

\pagestyle{plain}

Non-topological solitons \cite{tdlee}, in particular Q-balls
\cite{coleman1},  are known to exist in the limit of a large charge. 
Modern theories often envision the physics beyond the Standard Model (SM) 
as associated with a number of scalar fields with various charges.  
All supersymmetric generalizations of the SM, in particular the 
minimal model, MSSM, allow for a variety of baryonic and leptonic balls
built of squarks and sleptons, and the Higgs scalars \cite{ak_another}.  
These objects can be produced in the early Universe and can lead to
interesting  cosmological consequences.  Q-balls with a small charge can be
produced more easily at high temperatures.  It is of great interest,
therefore, to understand how small the charge of a stable Q-ball can be,
and what are the parameters of such solitons in the limit of small $Q$. 

The usual description of Q-balls \cite{coleman1} relies on the so-called
thin-wall approximation and is only valid for a very large charge $Q$. 
A naive extrapolation of these results beyond their domain of validity 
would lead one to conclude that no stable low-$Q$ solitons can exist. 
We will prove this not to be true, at least as long as the semiclassical
treatment of Q-balls remains appropriate.  There is no classical lower
limit on the charge of a stable Q-ball.  We will show that very small 
Q-balls (Q-beads) with charges  $Q \stackrel{>}{_{\scriptstyle \sim}} 1$ can 
exist and we will develop a formalism that yields an adequate description
of these solitons.

\section{Abelian Q-balls} 

Let us consider a field theory with a scalar potential $U(\vf) $ 
which has a global minimum $U(0)=0$ at $\vf=0$.
Let $U(\vf)$ have an unbroken global\footnote{
Q-balls associated with a local symmetry have been constructed 
\cite{l}.  An important qualitative difference is that, in the case of a
local symmetry, there is an upper limit on the charge of a stable $Q$
ball.} U(1) symmetry 
at the origin, $\vf=0$.  And let the scalar field $\vf$ have a unit charge
with respect to this $U(1)$.

The charge (taken to be positive for definiteness) of some field
configuration $\vf(x,t)$ is

\beq
Q= \frac{1}{2i} \int \vf^* \stackrel{\leftrightarrow}{\partial}_t  
\vf \, d^3x \ 
\label{Qt}
\eeq

Since a  trivial configuration $\vf(x)\equiv 0$ has zero charge, the
solution that minimizes the energy

\beq
E=\int d^3x \ \left [ \frac{1}{2} |\dot{\vf}|^2+
\frac{1}{2} |\nabla \vf|^2 
+U(\vf) \right]
\label{e}
\eeq
and has a given charge $Q>0$ must differ from zero in some (finite) domain. 
We will use the method of Lagrange multipliers to look for the minimum of 
$E$ at fixed $Q$.  We want to minimize 

\beq
\e = E+  \omega \left [ Q- \frac{1}{2i} \int \vf^* \dd_t \vf \, d^3x
\right ],   
\label{Ew1}
\eeq
where $ \omega$ is a Lagrange multiplier.  Variations of $\vf(x,t)$ and
those of $\omega$ can now be treated independently, the usual advantage of 
the Lagrange method.  

One can re-write equation (\ref{Ew1}) as 

\beq
\e = \int d^3x \, \frac{1}{2} \left | \dd_t \vf 
- i \omega  \vf 
\right |^2 \ + \ \int d^3x \, \left [\frac{1}{2}  |\nabla \vf |^2  
+ \hat{U}_\omega(\vf)
\right ] + \omega Q ,
\label{Ewt}
\eeq
where 

\beq
\hat{U}_\omega (\vf) = U(\vf)\ - \ \frac{1}{2} \, \omega^2 \, \vf^2. 
\label{Uhat}
\eeq

We are looking for a solution that extremizes $\e$, while all the physical
quantities, including the energy, $E$,  are time-independent.  Only the
first term in equation (\ref{Ewt}) appears to depend on time explicitly,
but it vanishes at the minimum.  To minimize this contribution to the energy,
one must choose, therefore,  

\beq
\vf (x,t) = e^{i\omega t} \vf (x),
\label{tsol}
\eeq
where $\vf(x)$ is real and independent of time.  For this solution,
equation (\ref{Qt}) yields 

\beq
Q= \omega \int \vf^2(x) \ d^3x
\label{Qw}
\eeq

It remains to find an extremum of the functional

\beq
\e = \int d^3x \, \left [\frac{1}{2} |\nabla \vf(x) |^2  
+ \hat{U}_\omega(\vf(x))
\right ] + \omega Q ,
\label{Ew}
\eeq
with respect to $\omega$ and the variations of $\vf(x)$ independently.
We can first minimize $\e$ for a fixed $\omega$, while varying the shape of
$\vf(x)$.  This, however, is identical to the problem of finding the bounce
$\bar{\vf}_\omega (x)$  for tunneling in 
$d=3$ Euclidean dimensions \cite{tunn0,tunn,linde} in the potential
$\hat{U}_\omega (\vf)$ (Fig. 1).  The first term in equation (\ref{Ew}) is
then nothing but the three-dimensional Euclidean action 
$S_3 [\bar{\vf}_\omega (x) ]$ of this bounce
solution. In what follows we will use this analogy 
extensively.  Over the years, the problem of tunneling has been studied
intensely, and many properties of $\bar{\vf}_\omega (x)$ and $S_3[
\bar{\vf}_\omega (x)]$ are well-known.

\begin{figure}
\setlength{\epsfxsize}{3.3in}
\centerline{\epsfbox{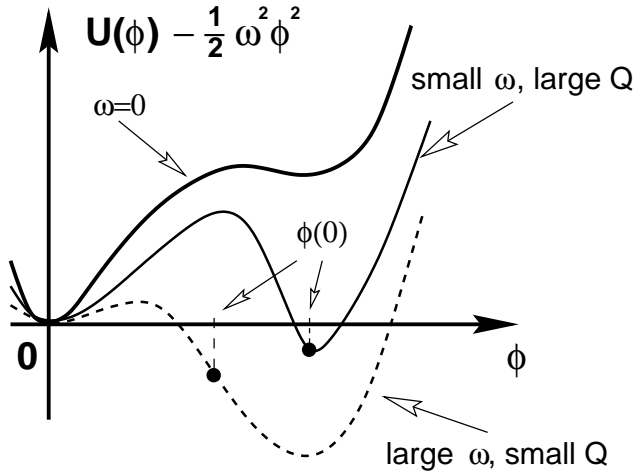}}
\caption{Finding a Q-ball is equivalent to finding a bounce that describes 
tunneling in the potential {$\hat{U}_\omega(\vf)= U(\vf)-(1/2) \omega^2
\vf^2 $}.  
Thin-wall approximation is good for large {$Q$} (thin solid
line), but breaks down when {$Q$} is small and, therefore, $\omega$ is
almost as large as the mass term at the origin.  In the latter case (dashed
line), the ``escape point'', {$\vf(0)$} is close to the zero of the
potential and is far from the global minimum.  This allows for an
alternative approximation that is good outside the thin-wall regime.  
}
\label{fig1}
\end{figure}

If the following condition  \cite{coleman1} is satisfied, 

\beq
U(\vf) \left/ \vf^2 \right. = {\rm min},
\ \ {\rm for} \ 
\vf=\vf_0>0 
\label{condmin}
\eeq
the corresponding effective potential $\hat{U}_{\omega_0}
(\vf)$, where $\omega_0=\sqrt{2 U(\vf_0^2)/\vf^2}$, will have two
degenerate minima, at $\vf=0$ and $\vf=\vf_0$.  

The existence of the bounce solution $\bar{\vf}_\omega (x)$ for 
$\omega_0<\omega < U''(0)$ 
follows \cite{tunn,cgm} from the fact that $\hat{U}_\omega (\vf)$ 
has a negative global minimum in addition to the local minimum at the
origin.  
From Ref. \cite{cgm} we know that the solution is
spherically symmetric: $\bar{\vf}(x)=\bar{\vf}(r), \ r=\sqrt{\vec{x}^2}$.

The soliton we want to construct is precisely this bounce for the right
choice of $\omega$, namely that which minimizes $\e$.
The last step is to find an extremum of 

\beq
\e= S_3 [\bar{\vf}_\omega (x)] + \omega Q 
\label{Ebounce}
\eeq
with respect to $\omega$.  The conditions for the existence of such an
extremum  will be discussed below.  
The values of $\omega$ range from $\omega_0$, the
minimal value for which $\hat{U}_\omega (\vf)$ is not everywhere positive
and, therefore, the bounce solution can exist, to $\omega = U''(0)$, at
which point $\hat{U}_\omega (\vf)$ has no barrier.  The latter does not
automatically mean that a non-trivial solution of the equations of
motion does not exist \cite{lw}.   However, for our purposes, we will only
have to consider $\omega $ in the range $\omega_0< \omega < U''(0)$. 

In the limit $\omega \rightarrow \omega_0+0$, the bounce (and, therefore, 
a Q-ball) solution can be analyzed in the thin-wall approximation
\cite{coleman1}.   However, for larger values of $\omega$, the thin-wall
approximation breaks down and so does the existence proof of
Ref. \cite{coleman1}. In our analysis, we will use a different
approximation \cite{linde}, that which is valid in the limit of large
$\omega$.  

Small, near-critical values of $\omega \approx \omega_0$, correspond to
the large values of charge $Q$.  There is a simple way to understand this. 
The first term in equation (\ref{Ebounce}), $S_3 [\bar{\vf}_\omega (x)]$, 
is a monotone decreasing function of $\omega$. (The smaller the barrier,
the more probable the tunneling is.  Thus $S_3$ must decrease with
$\omega$.) However, the last term in (\ref{Ew}),
$\omega Q$, increases with $\omega$.  Therefore, if there is a minimum, it
will be achieved for a smaller value of  $\omega$ if the $Q$ is larger, as 
illustrated in Fig. 2.  

\begin{figure}
\setlength{\epsfxsize}{3.3in}
\centerline{\epsfbox{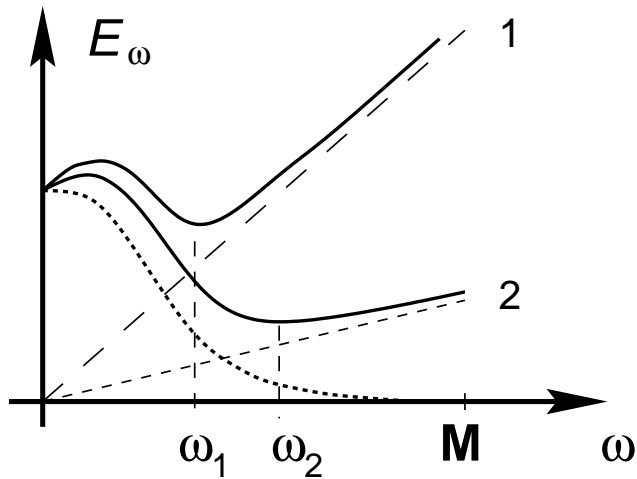}}
\caption{
Solitons with a smaller charge {$Q$} have larger values of {$\omega$}. 
The curved dotted line shows  {$S_3[\bar{\vf}_\omega (x)]$} as a function of
{$\omega$}.  The punctured straight lines 1 and 2 correspond to {$ 
\omega Q_1$} and {$ \omega Q_2$} for  {$Q_1>Q_2$}.  The minimum of
{$S_3[\bar{\vf}_\omega (x)] 
+\omega Q $} is reached at {$\omega=\ \omega_1$}, or {$\omega_2$},
respectively.  If {$Q_1>Q_2$}, then {$\omega_1<\omega_2$}. 
}
\label{fig2}
\end{figure}

If, however, $Q$ is sufficiently small, then the value of 
$\omega$ that minimizes $\e$ is large enough to destroy the near degeneracy
of the two minima (Fig. 1), and the thin-wall approximation breaks down. 

\section{Thick-wall approximation}

There is, however,  a powerful analytical
approximation\footnote{Another approach to calculating
$S_3[\bar{\vf}_\omega (x)]$ analytically outside the thin-wall
limit was proposed in Ref. \cite{kls}.  To calculate the bounce
action numerically, one can use the Improved Action method \cite{ak_n},
which is particularly useful for systems with many scalar degrees of
freedom. 
}
\cite{linde} that can be used for calculating
$S_3[\bar{\vf}_\omega (x)]$ 
in the  limit of very non-degenerate minima.  It is based on the
observation that for large $\omega$ (this case is analogous to tunneling
into a very deep minimum), the ``escape point'' $\bar{\vf}(0)$ (which is
also the maximal value of $\vf$ inside the Q-ball) is close to a zero of
$\hat{U}_\omega(\vf)$, and  is far from its minimum.   
As the barrier becomes smaller, the escape point $\bar{\vf}(0)$ moves closer 
to the origin (Fig. 1).  In this limit, one can neglect the dynamics at
large $\vf$ and retain only the quadratic and cubic terms in the potential. 

We will apply this strategy to minimize of $\e$ in equation 
(\ref{Ew}).
Let us consider a potential\footnote{
The $\vf^3$ term represents some $U(1)$-symmetric cubic interaction, {\it 
e.\,g.}, $(\vf^\dag  \vf)^{(3/2)}$.  In the MSSM, the requisite cubic
interactions arise from the tri-linear couplings of the
Higgs field to squarks and sleptons \cite{ak_another}.  For clarity, we
ignore the ``flavor structure'' of these cubic terms for now.}
$U(\vf) = \frac{1}{2} M^2 \vf^2 - A
\vf^3 + \lambda_4 \vf^4$, where $\vf$ has a unit charge with respect to 
some global $U(1)$ symmetry unbroken at $\vf=0$.  Then 

\beq
\hat{U}_\omega (\vf)= \frac{1}{2}(M^2-\omega^2) \vf^2 - A \vf^3 + 
\lambda_4 \vf^4 
\label{U234}
\eeq
The bounce for the potential $\hat{U}_\omega(\vf)$ for large $\omega$
(that is $0< (M-\omega)/(M+\omega) \ll 1$) is the same as that for
$\lambda_4 \rightarrow 0$.   As we show below, the condition for the
quartic terms to be negligible, $\lambda_4 \vf^4 \ll A \vf^3$ when $(1/2)
(M^2-\omega^2) \vf^2 \approx A\vf^3$, is satisfied whenever the charge is
small enough. One can, therefore, neglect the quartic term in
equation  (\ref{U234}) and introduce some new dimensionless variables 
for the space-time coordinates, $\xi_i= (M^2-\omega^2)^{1/2} x_i$, and for
the dynamical field, $\psi = \vf A/ (M^2-\omega^2)$.  In terms of these
variables, the expression for $\e$ becomes

\beq
\e= \frac{(M^2-\omega^2)^{3/2}}{A^2} \ \int d^3\xi \left [ \left (
\nabla_\xi \psi(\xi) \right)^2 +\frac{1}{2} \psi^2(\xi) - \psi^3(\xi) 
\right ] + w Q.
\label{sEw}
\eeq

The first term in equation (\ref{sEw}) is a dimensionful coefficient times
the action $S_\psi$ of the bounce in the potential $\Upsilon (\psi) = 
(1/2) \psi-\psi^3$.  It has been computed \cite{linde,bp} numerically: 
$S_\psi \approx 4.85$.  Its precise value is not important, we will only
need the fact that $S_\psi$ is independent of $M$, $\omega$, $A$, and $Q$.
The corresponding bounce $\bar{\psi}(\xi)$ has a radius $\sim 1$ in 
dimensionless units~\cite{linde}.  We have, therefore, extremized $\e$ with
respect to the variations of the $\psi(\xi)$, or, equivalently, those 
of $\vf(x)$.  What remains is to find a minimum of 

\beq
\e = S_\psi \frac{(M^2-\omega^2)^{3/2}}{A^2} + w Q
\label{wEw}
\eeq
with respect to $\omega$, $0 < \omega < M $.  This is possible as long as 

\beq
\epsilon \equiv \frac{Q A^2}{3 S_\psi M^2} < \frac{1}{2} 
\label{constr}  
\eeq
Minimum is achieved at $\omega = \omega_{min}= M \left [ (1+\sqrt{1-4
\epsilon^2})/ 2 \right ]^{1/2}$.  The resulting emergy (mass) of the
soliton can be expanded in powers of $\epsilon$: 

\beq
E=Q M \ \left [  1-\frac{1}{6} \epsilon^2 - \frac{1}{8} \epsilon^4 \ - \ 
O(\epsilon^6) 
\right ] 
\label{eQ}
\eeq
Since the mass of the soliton (\ref{eQ}) is less than $Q M$, it is stable
with respect to decay into the $\vf$ quanta.  The size $R$ of the soliton
is $\sim 1$ in dimensionless units $\sqrt{M^2-\omega^2} R$, that is 

\beq
R^{-1} \sim (M^2-\omega^2)^{1/2} \approx \epsilon \, M \, \left ( 
1 +\frac{1}{2} \epsilon^2 + \frac{7}{8} \epsilon^4 + O(\epsilon^6) 
\right )
\label{size}
\eeq

Two conditions must be satisfied in order for the approximation we just
used to be valid and self-consistent:   (i) the quartic terms must be
much smaller than the cubic and quadratic terms of the $\hat{U}$ in the 
vicinity of $\vf(0)$ (and, therefore, for every $\vf(x)$, because 
$\vf(0)=max [ \vf(x)]$); and (ii) $\e$ in equation (\ref{wEw}) must
have a minimum for $0<\omega<M$.  These two requirements set the following 
respective constraints on the charge $Q$.  

\beq
\left \{ 
\begin{array}{lll}
Q & \ll & 3 S_\psi M/\sqrt{\lambda_4} A \approx 
14.6 \, M/ \sqrt{\lambda_4} A \\  
& & \\ 
Q & < & 3 S_\psi M^2/ 2 A^2 \approx 7.28 \, M^2/A^2
\end{array} \right.
\label{Qconstr}
\eeq  
When  these  constraints are strongly violated, the thin-wall
approximation can be used.

We conclude that in the limit of small $Q$, the thin wall approximation must 
be replaced by the approximation (\ref{sEw}).  The existence of the solition
was proven in two steps.  First, we noted that $\e$ has an extremum with
respect to the variations of $\vf(x)$ for fixed $\omega$.  This relies on the
existence proof in Ref. \cite{tunn}. Second, using the ``thick-wall''
approximation, we showed that $\e$ has a minimum for some $\omega < M$. 

The expressions for
the energy (\ref{eQ}) and the size (\ref{size}) of the soliton are 
accurate when the cubic (tri-linear) couplings in the potential are
not too large, and when the charge is small (equatiuons (\ref{constr})
and (\ref{Qconstr})).   

There is no classical lower limit on charge: 
no matter how small  $Q$ is, there is a value of $\omega$ close to $M$, for
which the energy of the configuration is minimal.  However, for reasons of
quantum stability, $Q$ must be an integer.  Therefore, $Q\ge1$.  Also, 
as we show below, in the limit $Q\rightarrow 1$, the quantum corrections
can be significant.

\section{Classical stability and quantum corrections} 

We would like to examine the second variation of energy and prove the
stability of the soliton with respect to small variations that conserve
charge.  In the sector of a given charge $Q$ (the only subspace of the 
functional space in which we are interested), $E$ and $\e$ coincide, and so
do their variations. The time derivatives enter only in the first term
of equation (\ref{Ewt}), which is a non-negative function minimized by our
solution (\ref{tsol}).  What remains is to consider $(\delta^2 E)_{_Q}$ or  
$(\delta^2 \e)_{_Q}$ with respect to the variations of the time-independent
part of (\ref{tsol}), $\delta \vf(x)$.  For arbitrary $\delta \vf$, the
variation of $\omega $ is induced, so that the charge conservation
constraint (\ref{Qw}) is satisfied.  Starting with either of the
expressions (\ref{e}) or (\ref{Ew}),  one finds that 

\beq
(\delta^2 E)_{_Q} = \int \delta \vf 
[ -\Delta + U''(\bar{\vf}(x)) +3 \omega^2
] \delta \vf \ d^3x
\label{2var}
\eeq
where $3 \omega^2 $ is the contribution of the  second variation of the 
$(1/2) \omega^2 \vf^2$ term under the constant 
charge constraint: $\delta^2 \{ (1/2)
\omega^2 \vf^2 \}_{_Q} = 3 \omega^2 \delta \vf^2$.  If we expand $\delta
\vf =\sum c_i \psi_i$ in terms of the orthonormalized eigenvectors of the
differential operator in square brackets, $\psi_i$, corresponding to
the eigenvalues $\lambda_i$, then $(\delta^2 E)_{_Q} = \sum c_i^2
\lambda_i$.  Therefore, if the operator in square brackets in equation 
(\ref{2var}) has only positive eignevalues, then $(\delta^2 E)_{_Q}$ is
positive definite and the soliton is stable with respect to small
perturbations.  We, therefore, consider the following eigenvalue problem:

\beq
[-\Delta + U''(\bar{\vf_\omega}(x)) +3 \omega^2] \, \psi_i \ = \ \lambda_i
\psi_i 
\label{egeq}
\eeq
with the boundary condition $\psi(\infty)=0$ and the normalization
condition $\int \psi_i \psi_j = \delta_{ij}$. 



It is easy to see that for large enough $\omega$ operator (\ref{egeq}) has 
positive eigenvalues only.  Indeed, equation (\ref{egeq}) is just a
Schr\"odinger equation for the potential $U''(\bar{\vf}_\omega (x))+3
\omega^2$.  Since the potential $U(\vf)$ has a minimum at the origin, 
there exists a value $\vf_{con}$, such that $U''(\vf) >0$ for
$0<\vf<\vf_{con}$.  For large $\omega$, $\vf(0)$ is small (Fig. 1) and, for
a large enough $\omega$ (while still $\omega <U''(0)$),  $\vf(0)<\vf_{con}
$. Since $\forall x, \ \vf(x)<\vf(0)$, equation (\ref{egeq}) describes a 
quantum-mechanical  bound state of energy $\lambda_i$ in the potential that
is everywhere positive. Clearly, $\lambda_i$ is then also 
positive (and, in fact, $\lambda_i> 3\omega^2$).  

We have proven the stability of the Q-ball in the limit of large $\omega$ 
(small $Q$) with respect to small perturbations.  Coleman proved
\cite{coleman1} that Q-balls are stable with respect to all, not
only small, deformations in the limit of large $Q$.  We do not have a
rigorous proof of the soliton 
stability in the intermediate region, although it seems plausible.  

In order to evaluate the validity of the semiclassical approximation, 
one has to examine the magnitude of the quantum corrections to the mass of
the soliton. Semiclassical results are reliable if the quantum fluctuations
around the soliton are not large in comparison to its energy.  The
spectrum of such fluctuations is given by the same operator, $\delta^2
E/\delta \vf^2$, in the soliton background, renormalized with respect to the 
oscillations around the trivial vacuum solution.  The high-frequency modes
cancel out, but the low-frequency spectrum of $\delta^2 E/\delta \vf^2$ 
around the soliton will contain discrete levels which we would like to
estimate. 

Fortunately, we know something about the low-energy modes of a different
operator, which differs from (\ref{egeq}) by a constant.  Indeed, for the
three-dimensional bounce solution in the potential $\hat{U}_\omega (\vf)$, 

\beq
\frac{\delta^2 S_3}{\delta^2 \vf} = -\Delta + \hat{U}''_\omega
(\bar{\vf}_\omega(x)) =  
-\Delta + U''(\bar{\vf}_\omega(x)) - \omega^2. 
\label{tnn_opp}
\eeq 

The spectrum of the operator (\ref{tnn_opp}) was studied 
in Ref. \cite{tunn}.   We know that it has one negative eigenvalue 
$\tilde{\lambda}_1<0$.  However, we just proved that $\lambda_1=
\tilde{\lambda}_1 + 4 \omega^2 > 0$. 
Therefore, the lowest eigenvalue of the operator (\ref{egeq}),  
$\lambda_{1}$,  is  in the range 
$0<\lambda_1<4\omega^2$.  One can, therefore, expect the corrections
to the soliton mass squared to be (at the most) 
of order $\omega^2$, which, in the limit
of small $Q$, is $\sim U''(0)$.  These corrections will be small in
comparison to the soliton mass squared (\ref{eQ}) if $Q^2 \gg 1$.
However, the semiclassical approximation can become unreliable for 
calculating the masses of the solitons with charge $Q\sim 1$.
(At least, we don't have a proof to the contrary.) 
Nevertheless, even in this limit the size $R\sim \epsilon^{-1} M$ of
the soliton remains large in comaprison to its De Broglie wavelength, which
is an indication that semiclassical treatment may otherwise be appropriate
for $Q \sim 1$. 

\section{Virial theorem}

We will now prove a useful virial theorem which is valid for any
$Q$, and requires no approximation.  (In doing so, we will also illustrate
that the soliton is a global minimum with respect to the size variations.)
Let's consider a one-parameter family of functions obtained from the
solution $\bar{\vf}$ by expanding (contracting) it by a factor $\alpha$: 

\beq
\vf_\alpha(x) = \bar{\vf}(\alpha x)
\label{alpha}
\eeq
The energy of $\vf_\alpha$ is 

\beq
E_\alpha = \frac{1}{\alpha^3} \, 
\frac{Q^2}{ 2 \int  \bar{\vf}^2 d^3x} \ + \ 
\alpha \, T +  \alpha^3 \, V, 
\label{Ealpha} 
\eeq
where $T= \int \frac{1}{2} (\nabla \bar{\vf})^2 \ d^3x $
is the gradient energy, and $V=\int U(\bar{\vf}) d^3x $ is the potential
energy of the Q-ball, both positive.  

{\it Theorem:} 

\beq
T+ 3 V =  \frac{3\, Q^2}{ 2 \int  \bar{\vf}^2 \, d^3x} 
\label{virial}
\eeq

{\it Proof.} 
Since $\bar{\vf}$ is a stationary point of $E$ with respect to
all variations, it must, in particular, be an extremum with respect to
scaling (\ref{alpha}).  Therefore, $d E_\alpha/d\alpha =0$ at $\alpha=1$. 
This yields  relation (\ref{virial}).  $\Box$

If $Q=0$, there is no non-trivial solution, since $T\ge0$ and $V\ge
0$ would imply $T=V=0$.  Equation (\ref{virial}) shows, in particular, how 
the non-topological solitons evade the Derrick's theorem \cite{d}. 

We note that, since the coefficients of $1/\alpha^3$, $\alpha$,
and $\alpha^3$ in equation (\ref{Ealpha}) are all positive, the only
exteremum of $E_\alpha$ with respect to $\alpha$ (the size of the soliton)
is a global minimum.   

In summary, Q-balls with a small charge exist and are classically
stable.  They can be treated semiclassically at least as long as 
$Q^2 \gg 1$.  For $Q\approx 1$  the quantum corrections to the soliton 
mass are not necessarily small, but the Q-ball remains an extended object, 
whose size is large in comparison to its De Broglie wavelength. 

The author would like to thank M.~Shaposhnikov for many interesting
discussions, K.~Lee for a number of useful comments, 
and L.~\'Alvarez-Gaum\'e, S.~Dimopoulos, G.~Dvali and G.~Veneziano for
helpful conversations.


\begin{thebibliography}{99}

\bibitem{tdlee} T.~D.~Lee and Y.~Pang, Phys. Rep. 221 (1992) 251, and
references therein. 

\bibitem{coleman1} S.~Coleman, Nucl. Phys. {\bf B262} (1985) 263.

\bibitem{ak_another} A.~Kusenko, CERN-TH/97-70 (hep-ph/9704273).

\bibitem{l} K.~Lee, J.~A.~Stein-Schabes, R.~Watkins and L.~M.~Widrow,  
Phys. Rev. {\bf D39} (1989) 1665. 

\bibitem{tunn0} I.~Yu.~Kobzarev, L.~B.~Okun and M.~B.~Voloshin,
Yad.Fiz. {\bf 20} (1974) 1229 [Sov. J. Nucl. Phys. {\bf 20} (1975) 644]; 
P.~H.~Frampton, Phys. Rev. {\bf D15} (1977) 2922. 
 
\bibitem{tunn} S.~Coleman, Phys. Rev. {\bf D15}, 2929 (1977); C.G.~Callan 
and S.~Coleman, Phys. Rev. {\bf D16}, 1762 (1977). 

\bibitem{cgm} S.~Coleman, V.~Glaser and A.~Martin, Comm Math. Phys. {\bf 
58} (1978) 211.  

\bibitem{lw} K.~Lee and E.~Weinberg, Nucl. Phys. {\bf B267} (1986) 181. 

\bibitem{linde} A.~D.~Linde, Nucl. Phys. {\bf B216} (1983) 421. 

\bibitem{kls} A.~Kusenko, P.~Langacker and G.~Segr\`{e},  Phys. Rev. 
{\bf D54} (1996) 5824, Appendix B. 
 
\bibitem{ak_n} A.~Kusenko, Phys. Lett. {\bf B358} (1995) 51.

\bibitem{bp} E.~Brezin and G.~Parisi, J.~Stat.~Phys. {\bf 19} (1978) 269. 

\bibitem{d} G.~H.~Derrick, J. Math. Phys. {\bf 5} (1964) 1252.

\end{thebibliography}
\end{document}